\documentclass[longauth]{aa} 


\usepackage[varg]{txfonts}
\usepackage{microtype}
\usepackage{graphicx,color}

\usepackage[switch,modulo]{lineno}

\newcommand\arcdeg{\ensuremath{^{\circ}}}
\newcommand\arcs{\ensuremath{^{\prime\prime}}}
\newcommand\arcm{\ensuremath{^\prime}}
\newcommand\avg[1]{\ensuremath{\left<#1\right>}}
\newcommand\src{HESS\,J1018--589}
\newcommand\srcfermi{1FGL\,J1018.6--5856}

\begin{document}
\title{Discovery of variable VHE $\gamma$-ray emission from\\ the binary system \srcfermi}

\authorrunning{H.E.S.S.~Collaboration}
\titlerunning{Variable VHE emission from \srcfermi}

\author{H.E.S.S. Collaboration
\and A.~Abramowski \inst{1}
\and F.~Aharonian \inst{2,3,4}
\and F.~Ait Benkhali \inst{2}
\and A.G.~Akhperjanian \inst{5,4}
\and E.O.~Ang\"uner \inst{6}
\and M.~Backes \inst{7}
\and A.~Balzer \inst{8}
\and Y.~Becherini \inst{9}
\and J.~Becker Tjus \inst{10}
% ...
}

\institute{
Universit\"at Hamburg, Institut f\"ur Experimentalphysik, Luruper Chaussee 149, D 22761 Hamburg, Germany \and
Max-Planck-Institut f\"ur Kernphysik, P.O. Box 103980, D 69029 Heidelberg, Germany \and
Dublin Institute for Advanced Studies, 31 Fitzwilliam Place, Dublin 2, Ireland \and
National Academy of Sciences of the Republic of Armenia,  Marshall Baghramian Avenue, 24, 0019 Yerevan, Republic of Armenia  \and
Yerevan Physics Institute, 2 Alikhanian Brothers St., 375036 Yerevan, Armenia \and
Institut f\"ur Physik, Humboldt-Universit\"at zu Berlin, Newtonstr. 15, D 12489 Berlin, Germany \and
University of Namibia, Department of Physics, Private Bag 13301, Windhoek, Namibia \and
GRAPPA, Anton Pannekoek Institute for Astronomy, University of Amsterdam,  Science Park 904, 1098 XH Amsterdam, The Netherlands \and
Department of Physics and Electrical Engineering, Linnaeus University,  351 95 V\"axj\"o, Sweden \and
Institut f\"ur Theoretische Physik, Lehrstuhl IV: Weltraum und Astrophysik, Ruhr-Universit\"at Bochum, D 44780 Bochum, Germany
% ...
}

\offprints{\\
    E.~de O\~na Wilhelmi, \email{emma@mpi-hd.mpg.de}\\
    V.~Zabalza, \email{victor.zabalza@le.ac.uk}
}

\date{}
 
\abstract{
    Re-observations with the H.E.S.S.~telescope array of the
    very-high-energy (VHE) source \src\,A coincident with the {\em Fermi}-LAT
    $\gamma$-ray binary \srcfermi\ have resulted in a source detection significance of
    more than 9$\sigma$, and the detection of variability ($\chi^2$/$\nu$ of
    238.3/155) in the emitted $\gamma$-ray flux.  This variability confirms the
    association of \src\,A with the high-energy $\gamma$-ray binary detected by
    \emph{Fermi}-LAT, and also confirms the point-like source as a new
    very-high-energy binary system. The spectrum of \src\,A is best fit with a
    power-law function with photon index $\Gamma=2.20\pm0.14_{\rm
    stat}\pm0.2_{\rm sys}$.  Emission is detected up to $\sim$20\,TeV. The
    mean differential flux level is
    $(2.9\pm0.4)\times10^{-13}\,\mathrm{TeV^{-1}cm^{-2}s^{-1}}$ at 1 TeV,
    equivalent to $\sim$1\% of the flux from the Crab Nebula at the same energy.
    Variability is clearly detected in the night-by-night lightcurve.  When folded
    on the orbital period of 16.58 days, the rebinned lightcurve peaks in phase
    with the observed X-ray and high-energy phaseograms.  The fit of the
    H.E.S.S.~phaseogram to a constant flux provides evidence of periodicity at
    the level of $N_{\sigma}>3\sigma$. The shape of the VHE phaseogram and
    measured spectrum suggest a low inclination, low eccentricity system with a
    modest impact from VHE $\gamma$-ray absorption due to pair production
    ($\tau\lesssim1$ at 300 GeV). 
}

\maketitle

\section{Introduction} \label{sec1}

The region around the supernova remnant (SNR) SNR\,G284.3--1.8
\citep{1989PASAu...8..187M} shows two clearly distinct regions of
very-high-energy (VHE; E$>$100\,GeV) gamma-ray emission
\citep{2012A&A...541A...5H}; an extended emission named \src\,B likely
associated with the pulsar wind nebula (PWN) powered by the bright pulsar
PSR\,J1016--5857 \citep{2001ApJ...557L..51C,2004ApJ...616.1118C}, and the
point-like source HESS\,J1018--589\,A. The latter is positionally coincident
with \srcfermi, a point-like high-energy gamma-ray (HE; 100\,MeV$<$E$<$100\,GeV)
variable source detected by the \emph{Fermi} Large Area Telescope (LAT)
\citep{2010ApJS..188..405A}.

The $\gamma$-ray binary \srcfermi\ was detected in a blind search for periodic
sources in the {\em Fermi}-LAT survey of the Galactic Plane through the
modulation of its HE $\gamma$-ray flux \citep{2012Sci...335..189F}. Optical
observations show that the non-thermal source is positionally coincident with a
massive star of spectral type O6V((f)). The radio and X-ray flux from the source
are modulated with the same period of 16.58$\pm$0.02 days, interpreted as the
binary orbital period
\citep{2011ATel.3228....1P,2011ApJ...738L..31L,2012A&A...541A...5H,2013ApJ...775..135A}.

The spectrum of the periodic source in the {\em Fermi}-LAT domain exhibits a
break at $\sim$1\,GeV with best-fit values of $\Gamma{\rm_{HE}}$(0.1--1
GeV)=2.00$\pm$0.04 and $\Gamma{\rm_{HE}}$(1--10 GeV)=3.09$\pm$0.06 and an
integral energy flux above 100\,MeV of
(2.8$\pm$0.1)$\times$10$^{-10}$\,erg\,cm$^{-2}$\,s$^{-1}$. The HE $\gamma$-ray
spectral shape evolves with orbital phase, with a decrease in spectral curvature
at flux minimum of the emission (between phases 0.2 and 0.6) and a hardening of
the spectrum at flux maximum.

The best-fit position reported in the previous paper \citep{2012A&A...541A...5H}
for \src\,A, $\alpha$=10$^{\rm h}$18$^{\rm m}$59.3$^{\rm s}\pm$2.4$^{\rm s}_{\rm
stat}$ and $\delta$=--58\arcdeg56\arcm10\arcs$\pm$36$\arcs_{\rm stat}$
(J2000), is compatible with the 95\% confidence contour of 1FGL\,J1018.6--5856.
The VHE emission is well-described by a power-law function with a spectral index
of $\Gamma$=2.7$\pm$0.5$_{\rm stat}\pm$0.2$_{\rm sys}$, similar to the one
describing the VHE emission of the larger region \src\ B
($\Gamma$=2.9$\pm$0.4$_{\rm stat}\pm$0.2$_{\rm sys}$). No variability was found
in the H.E.S.S.~data set, although the contamination of the nearby source and
the uneven sampling of the observations prevented a firm conclusion at the time.

Here, a deeper study of HESS\,J1018--589\,A to assess its association with the
$\gamma$-ray binary is presented. In Section \ref{sec2}, the data sample and
results are described. In Section \ref{sec3}, the features of \src\,A are
discussed in light of the multi-wavelength observations available, and
conclusions are drawn in Section \ref{sec4}.

\section{Data Analysis and Results}
\label{sec2}

The H.E.S.S.~telescope array is a system of five VHE $\gamma$-ray imaging
atmospheric Cherenkov telescopes (IACTs) located in the Khomas Highland of
Namibia (23\arcdeg16\arcm18\arcs\ S, 16\arcdeg30\arcm00\arcs\ E). The fifth
telescope was added to the array in summer 2012 during the H.E.S.S.~phase-II
upgrade, increasing the energy coverage and boosting the system sensitivity. The
nominal sensitivity of the H.E.S.S.~phase-I array (excluding the large
telescope) reached in 25 hours is
$\sim$2.0$\times$10$^{-13}$~ph~cm$^{-2}$s$^{-1}$ (equivalent to 1$\%$ of the Crab
Nebula flux above 1\,TeV) for a point-like source detected at a significance of
5$\sigma$ at zenith. The stereoscopic approach results in a positional
reconstruction uncertainty of $\sim$6\arcm\ per event, good energy resolution
(15$\%$ on average) and an efficient background rejection
\citep{2006A&A...457..899A}.  H.E.S.S.-I observed the region towards the Carina
arm from 2004 to 2009. The data set presented in \cite{2012A&A...541A...5H}
was increased in the subsequent years from 40 hours to 63.3 hours effective time
with dedicated observations with the H.E.S.S.-I array to cover the orbital
phases in which the emission of \srcfermi\ observed in X-rays and HE increases.
The zenith angle at which the source was observed ranged from 35\arcdeg\ to
55\arcdeg\ resulting in a mean energy threshold of 0.35\,TeV. These new
observations were performed in wobble-mode, during which the telescopes were
pointed offset (0.7\arcdeg) from the nominal source location to allow
simultaneous background estimation. The data were analyzed using an improved
analysis technique (multivariate analysis, reaching $\sim$0.7\% of the Crab
Nebula flux at 1 TeV, 5 $\sigma$;
\citealt{2011APh....34..858B}) and cross-checked with the
Hillas second moment event reconstruction method \citep{2006A&A...457..899A} and
Model Analysis \citep{2009APh....32..231D}, including independent calibration of
pixel amplitudes and identification of problematic or dead pixels in the IACTs
cameras.  The spectra and light curves shown here are derived for a cut of 80
photoelectrons in the intensity of the recorded images. 
  
\begin{figure}
\centering
\includegraphics[width=\columnwidth]{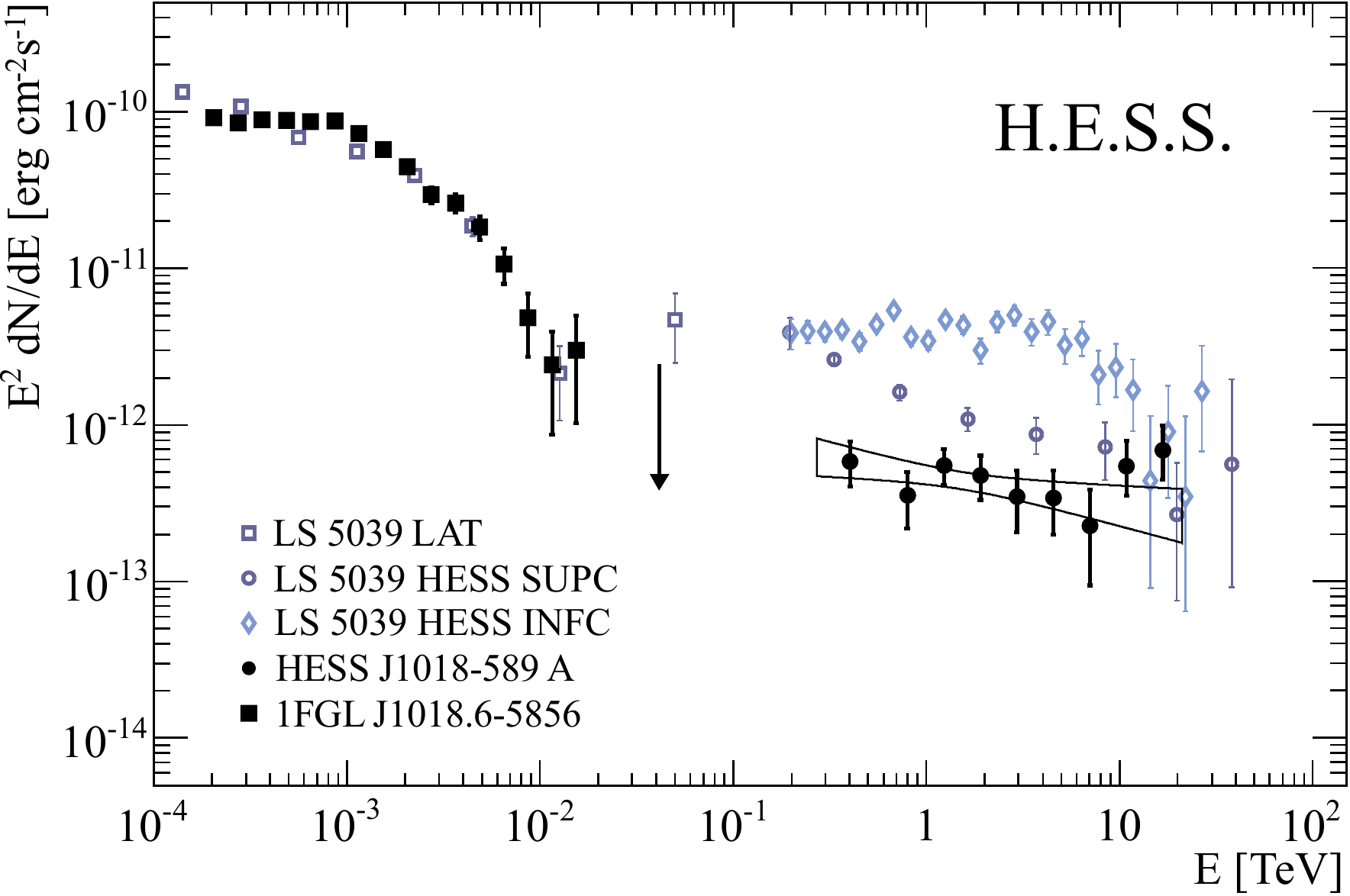}
\caption{The SED of \src\,A / \srcfermi\ is shown in black (filled
  squares and circles for the LAT and H.E.S.S.~detection
  respectively). For comparison, the SEDs of LS\,5039 during superior (SUPC) and
  inferior conjunction (INFC) are also included (blue points from
  \citealt{2012ApJ...749...54H,2005Sci...309..746A})}
\label{spec_total}
\end{figure}

The new analysis of \src\,A, using the larger data set, confirms the point-like
VHE $\gamma$-ray emission reported in \cite{2012A&A...541A...5H}. The
$\gamma$-ray signal is detected with a statistical significance of 9.3$\sigma$
pre-trials (derived using an oversampling radius of 0.10\arcdeg\ and
corresponding to more than 7.5$\sigma$ post-trials), centered at
$\alpha$=10$^{\rm h}$18$^{\rm m}$58$^{\rm s}\pm$5$^{\rm s}_{\rm stat}$ and
$\delta$=--58\arcdeg56\arcm43\arcs$\pm$30$\arcs_{\rm stat}$ (J2000). The
best-fit position is estimated by means of a maximum likelihood fit of the
exposure-corrected uncorrelated excess image. This position is compatible with
the position derived in \cite{2012A&A...541A...5H}, but the presence of the
nearby extended source \src\,B precludes from an improvement in the position
uncertainty even with the additional observation time.  The fitted extension is
compatible with the H.E.S.S.~point spread function (PSF, estimated to have a
mean 68\% containment radius of $\sim$0.1\arcdeg).  The obtained position is
used to derive the spectrum of the point-like source, integrating in a circle of
0.1\arcdeg\ around it and using a forward-folding maximum likelihood fit
\citep{2001A&A...374..895P}. The photon spectrum is well-described with a
power-law function with index $\Gamma$=2.20$\pm$0.14$_{\rm stat}\pm$0.2$_{\rm
sys}$ (Fig.~\ref{spec_total}) and the flux normalisation is
$N_0=(2.9\pm0.4_\mathrm{stat})\times10^{-13}\,\mathrm{TeV^{-1}cm^{-2}s^{-1}}$ at
1\,TeV. The systematic error on the normalisation constant $N_0$ is estimated to
be 20\% \citep{2006A&A...457..899A}. The better statistics allow for a better
determination of the spectral features of the point-like source compared with
the one presented in \cite{2012A&A...541A...5H}, including a clearer separation
from \src\ B. The nearby source introduces a maximum of 30\% contamination on
\src\,A, although above 1\,TeV, thanks to the better PSF, less than 10\%
contamination was calculated from a simultaneous fit of the two sources.

\begin{figure}
 \includegraphics{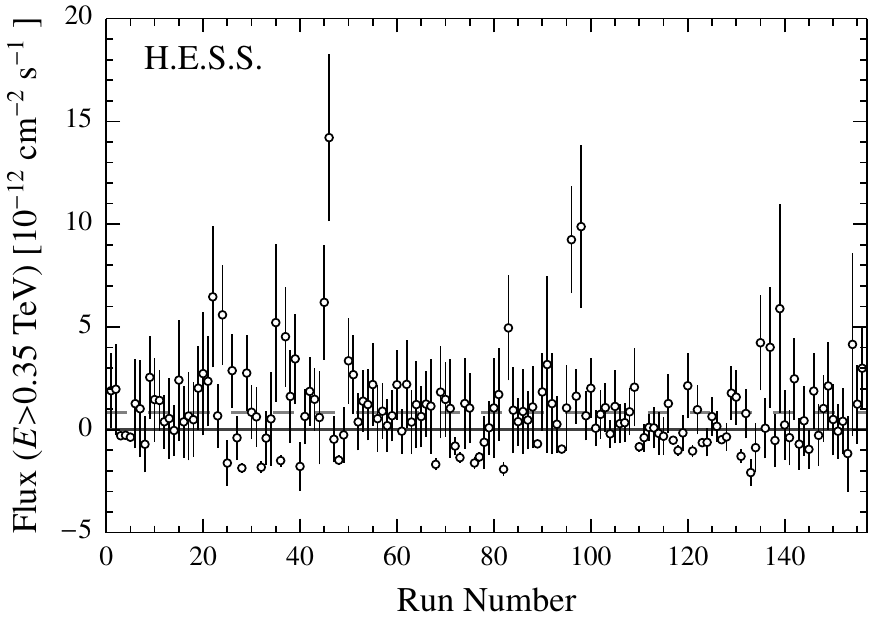}
 \caption{Lightcurve of the integral flux above 0.35\,TeV in a 0.1\arcdeg\ region
     centered on \src\,A binned by observation run, corresponding to
     approximately 30\,minutes of observation time per bin. 
     The dashed horizontal line shows the mean integral flux.}
     \label{LC}
\end{figure}

The light curve of the source above 0.35\,TeV, binned by observation run
(approximately 30\,minutes of observation time), is shown in Fig.~\ref{LC}. The
best-fit mean flux level above 0.35\,TeV is marked with a dashed gray line.
The lightcurve displays clear variability, with a $\chi^2$/$\nu$ of 238.3/155
(corresponding to 4.3$\sigma$) using a likelihood ratio test with a constant
flux as null hypothesis. 

\begin{figure}
\centering
\includegraphics{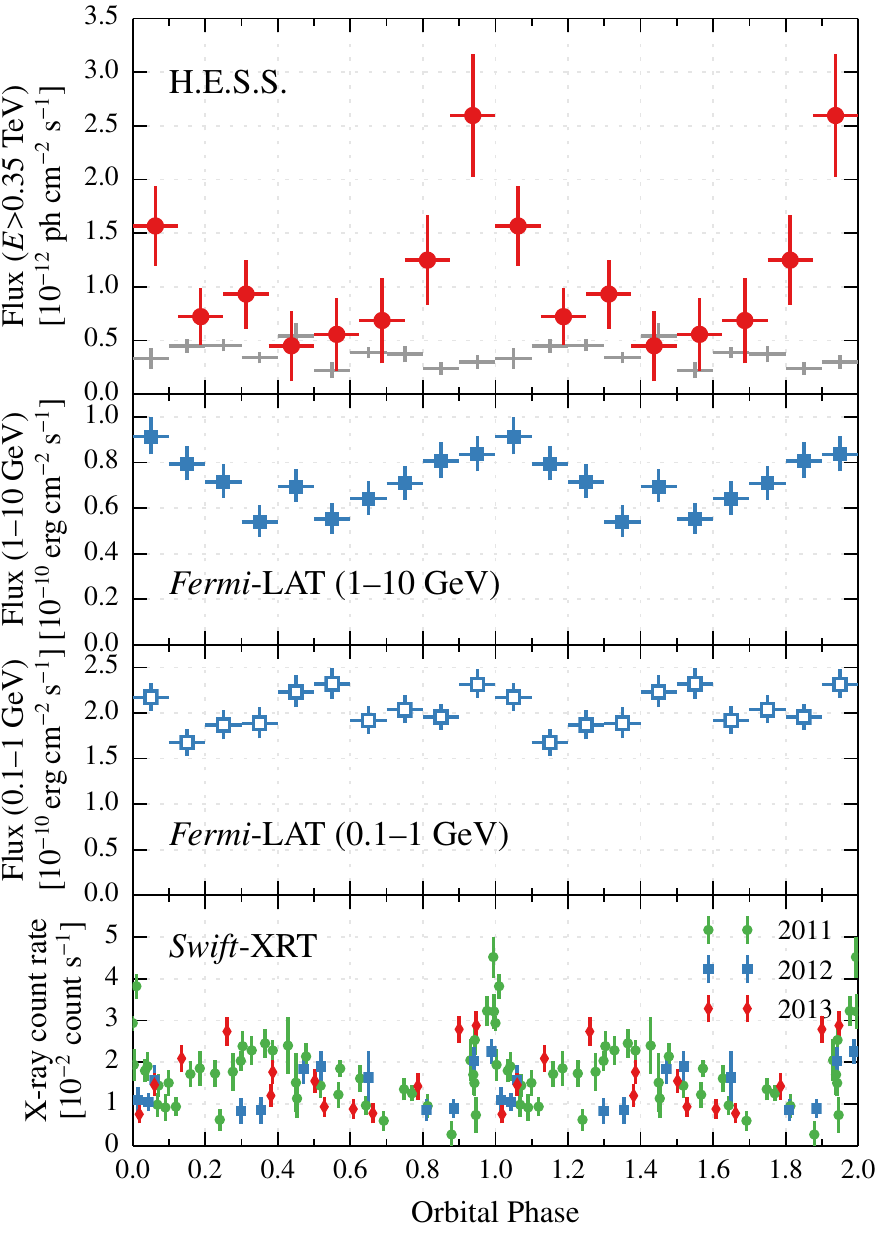}
\caption{
    VHE, HE, and X-ray fluxes of \srcfermi\ folded with the orbital period of
    P=16.58\,d. Two orbits are shown for clarity. \emph{Top}: VHE integral flux
    above 0.35\,TeV measure by H.E.S.S. (red circles). For comparison, a scaled
    lightcurve from the nearby bright source HESS\,J1023-589 is shown in gray.
    \emph{Middle top and middle bottom:} \emph{Fermi}-LAT lightcurve between 1
    and 10~GeV (solid blue squares) and between 0.1 and 1~GeV \citep[open blue
    squares;][]{2012Sci...335..189F}.  \emph{Bottom:} X-ray 0.3--10\,keV count
    rate lightcurve from 67 \emph{Swift}-XRT observations in 2011 (green), 2012
    (blue), and 2013 (red).
}
\label{Orbit}
\end{figure}

To investigate the periodicity of the source, the data were folded with the
16.58 day period found in the HE $\gamma$-ray observations (Fig.~\ref{Orbit},
top panel) using the reference time of T$_{\rm max}$=55403.3\,MJD as phase 0
\citep{2012Sci...335..189F} in a single trial. The number of bins in the
phaseogram was selected to obtain a significance of at least 1$\sigma$ in each
phase bin. For comparison, the same phaseogram is also shown for
HESS~J1023--589, a nearby bright $\gamma$-ray source expected to be constant.
The flux variation along the orbit shows a similar behaviour when comparing it
with the {\em Fermi}-LAT flux integrated between 1 and 10 GeV (Fig.~\ref{Orbit},
middle panel). An increase of the flux towards phase 0 is observed, with a
$\chi^2$/$\nu$ of 22.7/7 (3.1$\sigma$) when fitting the histogram to a constant
flux, providing evidence of periodicity at the \emph{a priori} selected period.
Unfortunately, the uneven sampling and large timespan of the observations did
not allow for an independent determination of the periodicity from the VHE
$\gamma$-ray data using a Lomb-Scargle test \citep{1982ApJ...263..835S}, since
the equivalent frequency is $\sim$8 times larger than the sample Nyquist
frequency. Finally spectral modulation was examined by deriving the photon
spectrum for observations in the 0.2 to 0.6 phase range (motivated by the {\em
Fermi}-LAT observations) and comparing it with the one derived at the maximum of
the emission in the complementary phase range. No spectral modulation was found
within the photon index errors ($\Delta\Gamma=0.36\pm0.43$) although it should
be noted that the data statistics in the 0.2 to 0.6 phase range are insufficient
(3$\sigma$ detection) to firmly conclude a lack of variation in the spectrum at
different orbital phases. 

In order to compare the VHE orbital modulation with the behaviour of the source
at X-ray energies, 67 \emph{Swift}-XRT observations of \srcfermi, performed
between 2011 and 2013 and with a median observation time of 2.2\,ksec,  were
analysed.  Early subsets of these observations were presented previously by
\cite{2012Sci...335..189F} and \cite{2013ApJ...775..135A}.  Cleaned event files
were obtained using \texttt{xrtpipeline} from HEAasoft v6.15.1. For each
observation, source count rates were extracted from a 1\,arcmin circular region
around the nominal position of \srcfermi, and background count rates extracted
from a nearby region of the same size devoid of sources. The resulting count
rate lightcurve, folded with the orbital period, is shown in the bottom panel of
Figure~\ref{Orbit}. The phaseogram displays a sharp peak around phase 0,
matching the location of the maximum in the VHE and HE phaseograms. There is an
additional sinusoidal component with a maximum around phase 0.3 and with lower
amplitude than the sharp peak at phase 0. 

\section{Discussion}
\label{sec3} 

The flux variability and periodical behaviour of \src\,A suggest the
identification of the VHE source with the $\gamma$-ray binary \srcfermi.  It
therefore becomes the fifth binary system, along with LS\,5039
\citep{2005Sci...309..746A}, LS\,I\,+61\,303 \citep{2006Sci...312.1771A},
PSR\,B1259--63 \citep{2005A&A...442....1A} and HESS\,J0632+057
\citep{2009ApJ...698L..94A} detected at VHE during multiple orbits, in addition
to the hint of a flaring episode from the X-ray binary Cygnus X-1
\citep{2007ApJ...665L..51A}.  When folded with the modulation period found at
other wavelengths, the rebinned VHE lightcurve shows a modulation, significant
at the 3.1$\sigma$ level, in phase with the HE $\gamma$-ray lightcurve.  The
phaseogram (Fig.~\ref{Orbit}) shows a similar behaviour (within the limited
statistics) to the high-energy lightcurve of \srcfermi\ showing a flux
increasing simultaneously to the one occurring in the HE and X-ray counterpart. 

Despite the different orbital behaviour at different wavelengths, the stars in
the $\gamma$-ray binaries \srcfermi\ and LS\,5039 are thought to be very
similar, with spectral types of O6V((f)) and O6.5V((f)), respectively
\citep{2012Sci...335..189F,2001A&A...376..476C}.  Unfortunately, the orbital
parameters of \src\,A are not yet known and only limited conclusions can be
drawn on the relation between the compact object and the massive star. Both
binary systems are composed of an almost identical massive star and a compact
object orbiting around it on a timescale of days. The four-times larger period
of \srcfermi\ implies a factor $\sim$2.5 larger semi-major axis than in
LS\,5039, and the low amplitude of the flux modulation observed by {\em
Fermi}-LAT, of the order of 25\%, can be interpreted as a sign of a
low-eccentricity orbit. Although the spectral index variability at HE
$\gamma$-ray is at odds with anisotropic IC being the only source of flux
variability, such low modulation amplitude would be difficult to realize under
the widely changing conditions of an eccentric orbit.  The behaviour of \src\,A
at different orbital phases is mimicked in X-rays, HE, and VHE, showing in all
cases a maximum flux near phase 0. There is a second sinusoidal component that
peaks at phase 0.3 and appears in radio \citep{2012Sci...335..189F} and X-rays
(Fig.~\ref{Orbit}, bottom panel), as well as a hint in the 0.1 to 1 GeV
\emph{Fermi}-LAT lightcurve peaking at phase 0.5 (Fig.~\ref{Orbit},
middle-bottom panel), but it is not observed (with the current statistics) at
higher energies. However, in LS\,5039 the VHE flux is correlated to the X-ray
flux but anti-correlated to the HE flux
\citep{2005Sci...309..746A,2009ApJ...697..592T,2009A&A...494L..37H,2012ApJ...749...54H}.
The HE spectral energy distributions (SEDs) of the two binary systems are
remarkably similar in shape and flux (see Fig.~\ref{spec_total}), although it
should be noted that the systems are believed to be located at different
distances: whereas LS\,5039 is $\sim$2.5 kpc away, \srcfermi\ is believed to be
located at 5$\pm$2 kpc, derived from the interstellar absorption lines of the
companion \citep{2012Sci...335..189F}. At VHE, LS\,5039 shows a clear spectral
modulation at different orbital phases, with a mean luminosity between 1 and 10
TeV of $\sim10^{33}(d_\mathrm{2.5\,kpc})^2 \mathrm{\,erg\,s^{-1}}$, similar to
the one found in \src\ A in the same energy range
($9.9\times10^{32}(d_\mathrm{5\,kpc})^2 \mathrm{\,erg\,s^{-1}}$).  However, the
ratios between the fluxes measured at HE and VHE of the two binary systems
differ substantially (see Figs.~\ref{spec_total} and \ref{Orbit}): whereas for
LS\,5039 the ratio between the fluxes at 1\,GeV and 1\,TeV varies between
$\sim$15 and 40 in superior and inferior conjunction, respectively, for
\srcfermi\ and \src\,A a ratio of $\sim$160 is found, with the TeV flux strongly
reduced with respect to the GeV flux when compared with LS\,5039.  Similar to
other binaries, the spectrum of \srcfermi\ measured at HE does not extrapolate
to VHE.

Regardless of the nature of the emission process responsible for the TeV
emission, the strong stellar photon field in the environment of the binary
system unavoidably leads to the absorption of $\gamma$-rays above $\sim$50\,GeV
through pair production
\citep{1994ApJS...92..567M,2005ApJ...634L..81B,2006A&A...451....9D}.  Assuming
that the VHE emission is due to anisotropic inverse-Compton (IC) scattering of a
leptonic population at a similar location as the HE one, the identical phasing
of HE and VHE would imply a low inclination of the orbit with respect to the
observing direction.
Furthermore, the sinusoidal modulation implies that the orbit cannot be highly
eccentric (or the density of photons would lead to strong variation). Under this
scenario, IC anisotropic emission would be most efficient when the emitter is
behind the star with respect to the emitter, i.e., at superior conjunction, and
therefore the flux maximum at phase 0 can be tentatively associated with this
orbital configuration.

\begin{figure}
    \includegraphics{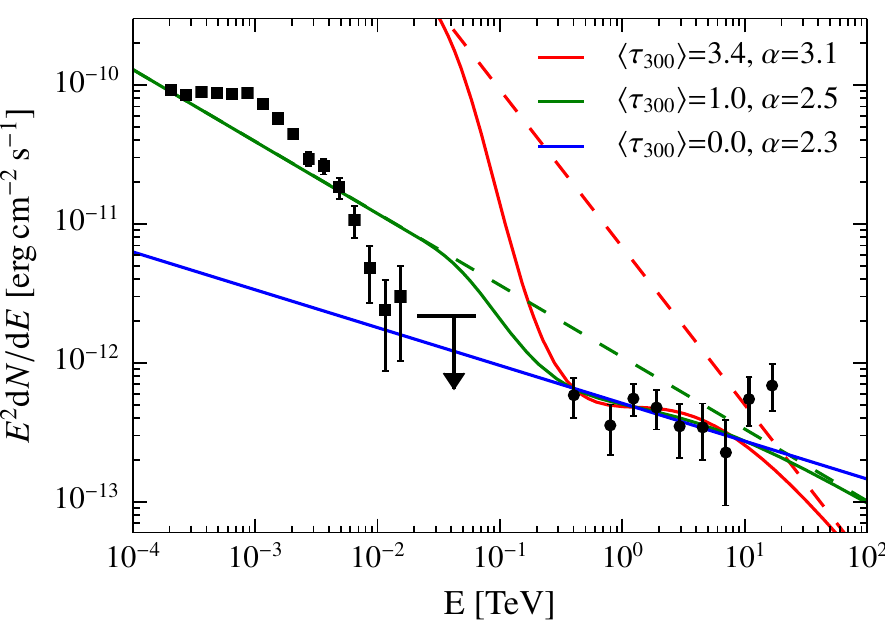}
    \caption{Results of the fit of a pair-production-absorbed power-law to the
        HE and VHE spectra of \srcfermi, with solid (dashed) lines indicating
        the absorbed (intrinsic) spectra. The emitter is assumed to be in the
        plane of the sky with respect to the star (i.e., on the plane of an
        orbit with $i=0\arcdeg$). The model with the maximum absorption
        compatible with the VHE spectral data is shown in red, the best-fit
        model with $\avg{\tau_{300}}=1$ in green, and the best-fit model with no
        absorption in blue.  Observational points are as in
        Fig.~\ref{spec_total}.\label{fig:tau-spec}}
\end{figure}

In order to illustrate the effects of absorption on the observed VHE spectrum,
in the following calculations a circular orbit on the plane of the sky
($i=0\arcdeg$) is assumed to exemplify the low, but likely non-zero, inclination
of the orbit, and take stellar parameters as in LS\,5039. A fit of a
pair-production-absorbed power-law function (of shape $\propto E^{-\alpha}
\exp(-\avg{\tau_E})$, where $\avg{\tau_E}$ is the energy-dependent
orbit-averaged optical depth and $\alpha$ the intrinsic spectral index) to the
measured VHE spectrum indicates that the maximum optical depth at 300\,GeV
compatible with the VHE data (at 68\% CL) is $\avg{\tau_{300}} \approx 3.4$,
with an intrinsic index of $\alpha\approx3.1$.  Figure~\ref{fig:tau-spec} shows
how the energy dependence of pair-production absorption results in a
power-law-like spectrum between 500\,GeV and 10\,TeV even for high optical
depths, as long as the intrinsic spectrum is steep enough.  This means that for
an optical depth of $\avg{\tau_{300}} \approx 3.4$, the steep spectral index
required to fit the VHE data would result in a large HE emission below 100~GeV,
where pair-production absorption is no longer significant, up to a factor 100
brighter than the flux observed by {\em Fermi}-LAT between 10 GeV and 100\,GeV.
Therefore, either the intrinsic emission from the VHE component has a sharp
spectral break between 80 and 200\,GeV, or the VHE instrinsic spectrum must be
significantly harder than $\alpha\approx3$.  Considering the latter, and taking
the {\em Fermi}-LAT flux between 10 GeV and 100\,GeV as an upper limit to the
emission of the VHE component at these energies, the optical depth should be
lower than 1, as illustrated by the green model in Fig.~\ref{fig:tau-spec}.  For
an orbital inclination of $i=0\arcdeg$, an optical depth lower than unity hints
towards an emitter located farther away from the star than the compact object
(at a distance of at least $\sim3\times10^{12}$\,cm from the compact object). At
higher orbital inclinations, the limit placed on the orbit-averaged optical
depth can not be directly related to the location of the emitter, given that the
optical depth would vary significantly along the orbit.  However, the
correlation between HE and VHE emission and the sharpness of the peak of VHE
emission at superior conjunction indicate that the optical depth at this
position must be low enough to not have a significant effect on the observed
flux modulation, therefore excluding an emitter close to the compact object for
high orbital inclinations.

Several mechanisms have been proposed to explain VHE variability and periodic
modulation via either IC processes or pion production of high energy protons
with the companion wind
\citep{1999APh....10...31K,2006A&A...459L..25B,2006ApJ...643.1081D,2006MNRAS.368..579B,2006A&A...451....9D,2008MNRAS.383..467K,2008APh....30..239S}.
In a leptonic scenario, the maximum energy of the H.E.S.S.~measured spectrum can
be used to derive further constraints on the location, magnetic field and
acceleration efficiency of the VHE emitter in HESS~J1018$-$589~A. Given the
energy of the stellar photons, IC scattering will take place in the deep
Klein-Nishina (KN) regime, in which all of the electron energy is transferred to
the scattered photons. In this scenario, the maximum energy detected (up to
$\sim$20\,TeV) would require the presence of ~20\,TeV electrons in the VHE
emitter, which in turn requires that they are accelerated faster than their
radiative energy loss timescale. The acceleration timescale can be expressed as: 
\begin{equation}
    t_\mathrm{acc}=\eta_\mathrm{acc} r_\mathrm{L}/c\approx 0.1 \eta_\mathrm{acc}E_\mathrm{TeV}
    B_\mathrm{G}^{-1}\,\mathrm{s},
\end{equation}
where $r_\mathrm{L}$ is the Larmor radius of the electron, $E_\mathrm{TeV}$ is
the electron energy in TeV units, $B_\mathrm{G}$ is the strength of the magnetic
field in Gauss, and $\eta_\mathrm{acc}>1$ is a parameter that characterizes the
efficiency of the acceleration (in general $\eta_\mathrm{acc}\gg1$, and only for
extreme accelerators does $\eta_\mathrm{acc}$ approach 1, i.e.~the Bohm
limit).  The balance between $t_\mathrm{acc}$ and the cooling time of electrons
in the KN regime, given by $t_\mathrm{KN}\approx10^3 d_{13}^2
E_\mathrm{TeV}^{0.7}\,\mathrm{s}$ \citep{2008MNRAS.383..467K}, where $d_{13}$ is
the distance to the optical star in units of $10^{13}$\,cm, implies
$E_\mathrm{max}\approx ( 10^4 B_\mathrm{G} \eta_\mathrm{acc}^{-1}
d_{13}^{-2})^{3.3}$\,TeV. For IC dominant losses, and considering the maximum
energy in the VHE spectrum, a minimum $B\gtrsim2.5\times10^{-4}\eta_\mathrm{acc}
d_{13}^2$\,G can be derived.  Furthermore, if non-radiative (adiabatic)
energy losses are negliglible, electron energy losses in the energy band
relevant to VHE emission would be dominated by the interplay between IC losses,
which in the KN regime decrease with energy, and synchrotron losses, which
increase with energy \citep{2005MNRAS.363..954M}.  For a power-law
$E_\mathrm{e}^{-p_\mathrm{inj}}$ injection spectrum with canonical
$p_\mathrm{inj}=2$, this results in a
hardening ($p_\mathrm{e}\sim1.3$) of the spectrum of the underlying steady-state
particle population up to the energy for which IC and synchrotron losses are
balanced, and a softening ($p_\mathrm{e}\sim3$) for higher energies \citep[see,
e.g.,][]{2005MNRAS.363..954M,2008A&A...477..691D}. The energy of the cooling
break, $E_\mathrm{break}$, can be found from the balance of IC and synchrotron
cooling timescales $t_\mathrm{KN}=t_\mathrm{syn}$, which, taking
$t_\mathrm{syn}\approx400 E_\mathrm{TeV}^{-1}B_\mathrm{G}^{-2}\,\mathrm{s}$,
results in $E_\mathrm{break}\approx0.58 (B_\mathrm{G}d_{13})^{-1.18}$\,TeV.  The
relatively hard VHE spectrum detected from \srcfermi\ requires an evolved
particle distribution with $p_\mathrm{e}\lesssim2$, indicating that $E_\mathrm{break}$
should be higher, or of the order of, the electron energies sampled by the TeV
spectrum. Considering $E_\mathrm{break}\gtrsim10$\,TeV, the magnetic field
strength is constrained by the VHE spectrum to $B\lesssim0.1 d_{13}^{-1}$\,G.

These constraints depend strongly on the location, acceleration efficiency and
magnetic field of the emission region.  An extended discussion of these
relationships for a VHE emitter in a binary system can be found in
\cite{2008MNRAS.383..467K}.  For the case of \srcfermi, the extension of a hard
VHE spectrum up to 20\,TeV indicates that acceleration/emission regions close to
the compact object require an extremely efficient acceleration process, with
$\eta_\mathrm{acc}\lesssim50$ and magnetic field strengths between 0.001 and
0.1\,G. If the emitter is located farther from the star, the constraint on the
acceleration efficiency is relaxed, but the upper limit on the magnetic field is
reduced to 0.03\,G at $d=3\times10^{13}$\,cm.  Regardless of the location of the
emitter, the requirement that the magnetic field strength is below 0.1\,G
indicates that, in this scenario, the particle population responsible for the
VHE emission would have a maximum 2--10\,keV X-ray flux of
$1.2\times10^{-14}\,\mathrm{erg\,cm^{-2}\,s^{-1}}$, more than an order of
magnitude lower than its detected X-ray flux of
$(6.5\pm0.7)\times10^{-13}\,\mathrm{erg\,cm^{-2}\,s^{-1}}$
\citep{2012A&A...541A...5H}, a similar situation to that found for LS~5039
\citep{2013A&A...551A..17Z}.

\section{Conclusions}
\label{sec4} 

The new observations of \src\,A with the H.E.S.S.~telescope array have increased
the significance of the detection up to $\sim$9$\sigma$, allowing the firm
identification of a new VHE binary system through the measurement of its
variable emission at a significance level of 4.3$\sigma$.
Folding the measured flux on a 16.58-day orbit results in a phaseogram similar
to the one observed at HE, with a wide peak around phase 0. The result of
fitting the phaseogram to a constant flux indicates evidence of periodic flux at
the 3.1$\sigma$ level. The phase-averaged photon spectrum extends up to
$\sim$20\,TeV, posing constraining limits on the magnetic field (0.001$<$B$<$0.1
G). Likewise, the spectral shape above 0.350 TeV limits the $\gamma$-$\gamma$
absorption and optical depth at 300 GeV to $\tau(300\,\mathrm{GeV})\lesssim1$,
which will be helpful in constraining the location of the emitter once the
orbital parameters are known.

Deeper observations with H.E.S.S.~II will improve the statistics at VHE and will
provide a measurement of the spectrum below $E<100$\,GeV, allowing the
investigation of key properties of the binary system such as spectral variation
within the orbit or the spectral shape at low energies.  Finally, the
investigation of the orbital parameters through radio and optical observations
is crucial to the understanding of the VHE emission mechanism in combination
with the periodicity and variability observed at lower energies.

\begin{acknowledgements}
The support of the Namibian authorities and of the University of Namibia in
facilitating the construction and operation of H.E.S.S.~is gratefully
acknowledged, as is the support by the German Ministry for Education and
Research (BMBF), the Max Planck Society, the German Research Foundation (DFG),
the French Ministry for Research, the CNRS-IN2P3 and the Astroparticle
Interdisciplinary Programme of the CNRS, the U.K. Science and Technology
Facilities Council (STFC), the IPNP of the Charles University, the Czech
Science Foundation, the Polish Ministry of Science and  Higher Education, the
South African Department of Science and Technology and National Research
Foundation, and by the University of Namibia. We appreciate the excellent work
of the technical support staff in Berlin, Durham, Hamburg, Heidelberg,
Palaiseau, Paris, Saclay, and in Namibia in the construction and operation of
the equipment.
\end{acknowledgements}

\bibliographystyle{aa}

\end{document}